\documentclass[final,3p,times]{elsarticle}

\usepackage{amssymb}
\usepackage{listings}
\usepackage{color}
\usepackage{verbatim}

\biboptions{}
\newcommand{\dint}{{\rm d}}
\newcommand{\sartre}{Sar{\it t}re}
\newcommand{\xpom}{x_{I\!\!P}}

\journal{Computational Physics Community}

\begin{document}

\begin{frontmatter}



\begin{flushright}
Draft Version 5, \today 
\end{flushright}

\title{The dipole model Monte Carlo generator \sartre~1}
\author{Tobias Toll\corref{cor1}\fnref{label2}}
\ead{ttoll@bnl.gov}
\author{Thomas Ullrich \corref{cor1}\fnref{label2}}
\ead{thomas.ullrich@bnl.gov}
\address{Brookhaven National Laboratory, Upton, NY}



\author{}

\address{}

\begin{abstract}
We present the Monte Carlo generator \sartre~ for simulating 
diffractive exclusive vector meson production and DVCS in
electron-proton, electron-ion, and hadron-hadron collisions.
\end{abstract}


\end{frontmatter}


\section*{Program Summary}
{\it Program Title:} \sartre~1.0

{\it Authors:} Tobias Toll, Thomas Ullrich

{\it Licensing provisions:} GPL version 2

{\it Programming language:} C/C++

{\it Computer for which the program is designed and others on which it is operable:}
Any with standard C/C++ compiler. Tested on Linux and MacOS.

{\it Separate documentation available via:} \verb+https://code.google.com/p/sartre-mc/+


{\it Nature of physical problem:} 
Simulate diffractive exclusive vector meson and deeply virtual Compton scattering (DVCS) production in electron-nucleus scattering
where the exchanged virtual photon interacts
coherently with a large region of the nucleus. To calculate the cross section correctly it has to be 
averaged over all possible configurations of nucleon positions within the nucleus.

{\it Method of solution:} To make an arithmetic average of the quantum mechanical amplitude over nucleon configurations numerically and store the result in look-up tables.

{\it Implemented processes:} The following processes can be simulated:
\begin{eqnarray}
  e+p\rightarrow e' + V + p' \nonumber \\
  e+A\rightarrow e' + V + A' \nonumber \\
  p+p\rightarrow p' + V + p' \nonumber \\
  p+A\rightarrow p' + V + A' \nonumber \\
  A+A\rightarrow A' + V + A' \nonumber \\
\end{eqnarray}
where $V$ is a $J/\psi$, $\phi$, or $\rho$ vector meson, or a real photon (DVCS). All processes are
mediated by a virtual photon and a pomeron.

The present version is applicable for these processes at future electron-hadron colliders, such as the EIC and the LHeC,
as well as HERA, RHIC, and the LHC.

{\it Restrictions to physics problem:} The program is reliable for process at $\xpom< 10^{-2}$, and large $\beta=x/\xpom$.

{\it Other Programs used:} ROOT and GSL for numeric algorithms and other various tasks throughout the program.  BOOST for multi-threaded integration (optional), GEMINI++ for nuclear break-up and CUBA for multidimensional numerical integration (the latter two supplied with the program package).
Uses cmake for building and installing.

{\it Unusual feature:} None

{\it Running time:} On a MacBook Pro with a 2.66 GHz Intel Core i7 processor, 
event generation takes $\sim 0.1$ ms/event without correction and nuclear breakup, $\sim 0.2$ ms/event with the recommended corrections 
switched on, and $\sim 6$ ms/event when running with corrections and nuclear breakup.

{\it Download of the program:} \verb+http://code.google.com/p/sartre-mc/+

\section*{LONG WRITE-UP}
\section{Introduction}
The dipole model has been very successful in describing
exclusive diffraction at HERA, and is presently the most common approach at small $x$.
It describes the interaction between the electron and the proton by letting the virtual photon
split into a quark-anti-quark pair, which forms a color dipole. This dipole subsequently interacts
with the proton in the proton's rest frame. The dipole model in its present form was suggested by 
Golec-Biernat and W\"usthoff (GBW) \cite{GolecBiernat:1998js, GolecBiernat:1999qd},
who observed that a simple ansatz of the dipole model, integrated over impact parameter,
was able to simultaneously describe the total inclusive {\it and} diffractive cross sections,
the latter of which the collinear DGLAP formalism underestimates severely. The GBW model 
contains saturation in the small $x$ regime in a natural way. However, the GBW model fails at
describing the high $Q^2$ scaling violation observed in the inclusive cross section measured at HERA,
something the DGLAP formalism is able to describe perfectly. 
Bartels, Golec-Biernat, and Kowalski (BGBK) therefore included an explicit DGLAP gluon distribution
into the dipole formalism \cite{Bartels:2002cj}, taken at a scale
directly linked to the dipole size. The BGBK model replicates the GBW
model where it is applicable and also manages to describe the $Q^2$
dependence of the cross sections. However, this approach still
integrates out the impact parameter dependence of the interaction,
without which the $t$-dependence of the cross section is undetermined.  The
impact parameter dependence was introduced in the dipole model by
Kowalski and Teaney \cite{hep-ph/0304189} and then modified to also
include exclusive processes by Kowalski, Teaney, and Motyka
\cite{Kowalski:2006hc}. This dipole model goes by the name bSat (or
sometimes IPSat). Kowalski and Teaney also introduced a linearized version
of bSat, called bNonSat, in order to separate and thus isolate
non-linear effects to the cross sections \cite{hep-ph/0304189}.

In a recent paper we described in detail how the bSat and bNonSat models can be extended
to also describe DIS with heavy nuclei \cite{sartre}. We have implemented the
bSat and bNonSat dipole models, for both protons and nuclei, into a Monte Carlo event generator named
\sartre, which is the focus of this paper.

The main motivation for creating \sartre~comes as a part of the effort to realize a future
electron-ion collider (EIC) \cite{Deshpande:2012bu}. 
While the legacy of HERA is a plethora of physics generators describing
all aspects of electron-hadron collisions ({\it e.g.}
PYTHIA6 \cite{Sjostrand:2000wi}, HERWIG++ \cite{Bahr:2008pv}, LEPTO
\cite{Ingelman:1996mq}, PEPSI \cite{Mankiewicz:1991dp}, RAPGAP
\cite{Jung:1993gf}, ARIADNE \cite{Lonnblad:1992tz}, CASCADE
\cite{Jung:2001hx}, SHERPA \cite{Gleisberg:2008ta}), there is a dearth of such generators describing
electron-ion collisions. The only exception known to us is DPMJET-III \cite{Roesler:2000he}, which
however is limited to photo-production ($Q^2=0$) and high-mass diffractive dissociation including 
multiple jet production. 
One of the key measurements at an EIC is that of the {\it spatial} distribution of gluons at small $x$,
which has never been studied experimentally. 
To obtain the spatial gluon distribution, one measures the diffractive cross section,
$d\sigma/dt$, at low-$x$ over a large range of $t$. A non-trivial Fourier transform from momentum space
to coordinate space provides then the desired source distribution $F(b_T)$. 
However, a direct measurement of $t = (\textbf{p}-\mathbf{p}^\prime)^2$ is not possible in $eA$ collisions since the scattered
ion ($\mathbf{p}^\prime$) is not sufficiently well separated from the beam. 
The only processes that allow access to $t$ with sufficiently high precision
are exclusive diffractive processes, such as exclusive vector-meson production and
deeply virtual Compton scattering (DVCS). Here $t$ can be calculated from the measured vector meson (or photon), the scattered electron, and the known beam energies.
From an experimental standpoint this is rather challenging measurement that requires a carefully designed detector, making
detailed simulations of the underlying physics an imperative. 

There is presently one class of events at existing hadron collider experiments (RHIC and LHC)
that can be described by \sartre: those of ultraperipheral collisions (UPC) between hadrons. In 
these events, the impact parameter between the colliding hadrons is so large that the long-range
electro-magnetic force dominates over the short-ranged strong force. Therefore, these
interactions are mediated by a virtual photon, and can thus potentially be described by the
dipole model.

The program is named after the existentialist philosopher Jean-Paul Sartre. 
According to existentialism, existence comes before essence, but Sartre
mentions the following important exception to that rule
in a lecture held 1945 \cite{ExistentialismHumanism}:
\begin{quote}
If one considers an article of manufacture as, for example
[...] a paper-knife -- one sees that 
it has been made by an artisan who had a conception of it; and he has paid attention, equally, to the 
conception of a paper-knife and to the pre-existent technique of production which is a part of that 
conception.
[...]Thus the paper-knife is at the same time an article producible in 
a certain manner and one which, on the other hand, 
[...] serves a definite purpose, for one cannot suppose 
that a man would produce a paper-knife without knowing what it was for. 
Let us say, then, of the 
paperknife that its essence 
[...] precedes its existence. 
[...]Here, then, we are viewing the world from a technical 
standpoint, and we can say that production precedes existence.
\end{quote}
The purpose of \sartre~is to provide simulations of events at an electron-ion collider, not yet in
existence. Therefore, one may say, \sartre~provides the artisan's pre-conception of the essence
of an EIC, which is a necessary guidance for the construction of the machine and its detectors.

The paper is organized as follows: In section \ref{dipolemodel} we give a description
of the dipole models implemented in \sartre. In section \ref{program}, the program is given
an overview description. Finally, in section \ref{examples}, we give an example of a
user program running \sartre.
\section{The dipole model in $ep$ and $e$A}
\label{dipolemodel}
\begin{figure}
   \begin{center}
    \includegraphics[width=0.5\paperwidth]{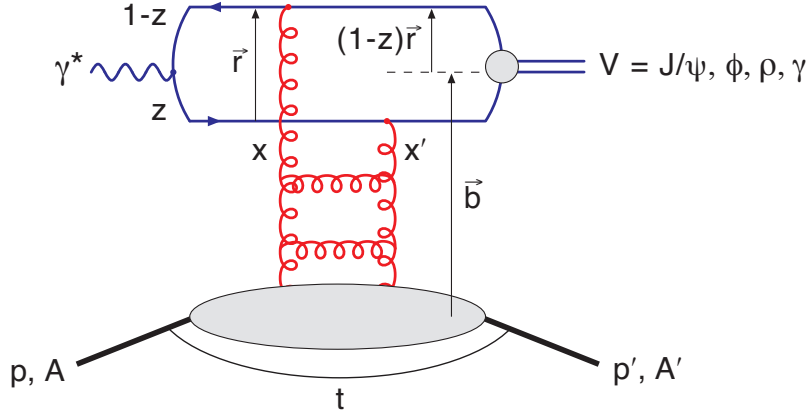}
    \caption{\label{fig:dipole}A schematic view of the dipole model and its variables. See text
    	for details.}
    \end{center}
\end{figure}
The amplitude for producing an exclusive vector meson or a real photon
diffractively in an interaction between a virtual photon and a proton can be written as:
\begin{eqnarray}
  \mathcal{A}_{T,L}^{\gamma^*p}(x_{I\!\!P},Q,\Delta) =
  i\int{\rm d} r\int\frac{{\rm d} z}{4\pi}\int{\rm d}^2{\bf b}\left[\left(\Psi^*_V\Psi\right)(r, z)\right]
2\pi rJ_0([1-z]r\Delta) 
  e^{-i{\bf b}\cdot{\Delta}}\frac{{\rm d}\sigma_{q\bar q}^{(p)}}{{\rm d}^2{\bf b}}(x_{I\!\!P}, r, {\bf b}),
  ~~&~&
    \label{eq:ttepamplitude}
\end{eqnarray}
where $r$ is the dipole's size, ${\bf b}$ its impact parameter
in relation to the proton's mass center, and $z$ is the momentum fraction of the dipole
taken by the quark. The variable $x_{I\!\!P}$ is the pomeron's momentum fraction
of the proton, $Q^2$  the virtuality of the photon, and $\Delta$ the virtuality of the .
The virtual photon is either longitudinally or transversely polarized, denoted
by $L$ and $T$, respectively. $J_0$ is a Bessel function and $\left(\Psi^*_V\Psi\right)(r, z)$ is 
the wave-function overlap between the incoming virtual photon and the outgoing vector-meson 
or real photon.

The proton dipole cross section is given by:
\begin{eqnarray}
	\frac{ \dint_{q\bar q}^{(p)} }{ \dint^2{\bf b} }(x_{I\!\!P}, r, {\bf b})=2\mathcal{N}^{(p)}(x_{I\!\!P}, r, {\bf b})
		=2\left[1-\Re(S)\right]
	\label{eq:epcs}
\end{eqnarray}
where $\mathcal{N}^{(p)}$ is the scattering amplitude of the proton. We only use the real part
of the $S$-matrix. In the bSat model the scattering amplitude is:
\begin{eqnarray}
	\mathcal{N}^{(p)}=1-\exp\left(-\frac{\pi^2}{2N_C}r^2\alpha_{\rm S}(\mu^2)x_{I\!\!P}g(x_{I\!\!P}, \mu^2)T(b)\right)
	\label{eq:epsa}
\end{eqnarray}
where $\mu^2=4/r^2+\mu_0^2$ and $\mu_0^2$ is a cut-off scale in the DGLAP evolution of the gluons 
$xg(x, \mu^2)=A_gx^{-\lambda_g}(1-x)^{5.6}$. 
The impact parameter dependence is introduced through the proton's profile function
$T(b)=1/(2\pi B_G)\exp(-b^2/(2B_G))$. All parameter values are
determined through fits to HERA data \cite{Kowalski:2006hc}, and are found to be 
$B_G=4~{\rm GeV}^{-2}$, $\mu_0^2=1.17$
GeV$^2$, $\lambda_g=0.02$, and $A_g=2.55$. Also, the four lightest
quark masses are treated as parameters in the model, and are taken to
be: $m_u=m_d=m_s=0.14$ GeV, $m_c=1.4$ GeV.

The nuclear scattering amplitude is constructed from that of the proton:
\begin{eqnarray}
	1-\mathcal{N}^{(A)}(x_{I\!\!P}, {\bf r}, {\bf b})=
	\prod_{i=1}^A
    \left(1-\mathcal{N}^{(p)}(x_{I\!\!P}, {\bf r}, |{\bf b}-{\bf b}_i|)\right)
    \label{eq:eptoea}
\end{eqnarray}
where ${\bf b}_i$ is the position of each nucleon in the nucleus in the transverse plane. We distribute
the nucleons according to the Woods-Saxon function projected onto the transverse plane.
Combining equations (\ref{eq:epcs}), (\ref{eq:epsa}), and (\ref{eq:eptoea}), the bSat
dipole cross section for $\gamma^*$A becomes:
\begin{eqnarray}
	\frac{1}{2}	
  \frac{{\rm d}\sigma_{q\bar q}^{(A)}}{{\rm d}^2{\bf b}}(x_{I\!\!P}, r, {\bf b}, \Omega_j)=
  1-\exp\bigg(-\frac{\pi^2}{2N_C}r^2    
    \alpha_S(\mu^2)x_{I\!\!P}g(x_{I\!\!P},\mu^2) 
    \sum_{i=1}^AT(|{\bf b}-{\bf b}_i|)\bigg)
  \label{eq:bSateA}
\end{eqnarray}
where $\Omega_j=\{{\bf b}_1, {\bf b}_2, \dots, {\bf b}_A\}$ represents a specific Woods-Saxon configuration of nucleons.

In $ep$ the diffractive cross section is given by the absolute square of the amplitude:
\begin{eqnarray}
	\frac{\dint\sigma^{\gamma^*p}_{T, L} }{\dint t}=\frac{1}{16\pi}\left|\mathcal{A}^{\gamma^*p}(x_{I\!\!P}, Q^2, t)\right|^2.
\end{eqnarray}

In $e$A one has to average the squared amplitude over all possible nucleon configurations $\Omega$:
\begin{eqnarray}
	\frac{\dint\sigma^{\gamma^*A}_{T, L} }{\dint t}=
	\frac{1}{16\pi}\left<\left|\mathcal{A}^{\gamma^*p}(x_{I\!\!P}, Q^2, t)\right|^2\right>_\Omega.
\end{eqnarray}

In $e$A there are two possible scenarios: either the interaction between the nucleus and the dipole
is elastic, or it is inelastic, in which case the nucleus de-excites subsequent to the interaction
by breaking up into color-neutral fragments. The first case is called coherent and the latter is called
incoherent. According to the Good-Walker picture, the incoherent cross section is 
proportional to the variance of the amplitude:
\begin{eqnarray}
    \frac{{\rm d}\sigma_{\rm incoherent}}{{\rm d} t}=
    \frac{1}{16\pi}\bigg(\left<\left|\mathcal{A}(x_{I\!\!P}, Q^2, t, \Omega)\right|^2\right>_\Omega -
	\left|\left<\mathcal{A}(x_{I\!\!P}, Q^2, t, \Omega)\right>_\Omega\right|^2\bigg) 
\end{eqnarray}
where the second term on the R.H.S.~is the coherent part of the cross section.
To calculate the diffractive cross sections for incoherent and coherent interaction
therefore becomes a matter of calculating the second and first moments of the
amplitude respectively.

For the first
moment there is a closed expression for the average of the
dipole cross section \cite{hep-ph/0304189}:
\begin{eqnarray}
  \left<\frac{{\rm d}\sigma_{q\bar q}}{{\rm d}^2{\bf b}}\right>_\Omega = 
  2\left[1-\left(1-\frac{T_A({\bf b})}{2}\sigma^p_{q\bar q}\right)^A\right]
  \label{eq:analytical}
\end{eqnarray}
where $\sigma^p_{q\bar q}$ is the $ep$ dipole cross section, eq.~(\ref{eq:epcs}),
integrated over the impact parameter, and $T_A$ is the spatial density profile of 
ions which is taken to be the Woods-Saxon potential in transverse space. 

For the second moment of the amplitude, no analytical expression
exists.  We define the average of an observable 
$\mathcal{O}(\Omega)$ over nucleon configurations $\Omega_i$ by:
\begin{eqnarray}
    \left<\mathcal{O}\right>_\Omega=\frac{1}{C_{\rm max}}\sum_{i=1}^{C_{\rm max}}\mathcal{O}(\Omega_i).
    \label{eq:average}
\end{eqnarray}
For a large enough number of configurations $C_{\rm max}$ the sum on
the R.H.S.  will converge to the true average. For the total
diffractive cross section one gets:
\begin{eqnarray}
    \frac{{\rm d}\sigma^{\gamma^*{\rm A}}}{{\rm d}t}(x_{I\!\!P}, Q^2, t) = 
    \frac{1}{16\pi}\frac{1}{C_{\rm max}}\sum_{i=1}^{C_{\rm max}}\left|\mathcal{A}(x_{I\!\!P}, Q^2, t, \Omega_i)\right|^2.~~~~
\end{eqnarray}
For large $|t|$, the variance of the amplitude is several orders of magnitude larger than
the average.  This means that the convergence of the sum in
eq.~(\ref{eq:average}) becomes extremely slow. For the first moment 
we therefore use eq.~(\ref{eq:analytical}). For the second moment
we have shown in \cite{sartre} that 500 configurations give a good
convergence.

\subsection{The non-saturated dipole model}
In order to separate saturation from other small-$x$ effects, 
a linearized version of the dipole model, called the bNonSat model 
\cite{hep-ph/0304189}, is implemented in \sartre.
It is obtained by linearizing the dipole cross section of the bSat model. By doing so
the gluon density becomes unsaturated for small $x$ and for large
dipole sizes $r$.

In the proton case, the bNonSat dipole cross section is obtained by
keeping the first term in the expansion of the exponent in the
bSat dipole cross section \cite{hep-ph/0304189}:
\begin{eqnarray}
  \frac{{\rm d}\sigma_{q\bar q}^{(p)}}{{\rm d}^2b}
  =\frac{\pi^2}{N_C}r^2\alpha_s(\mu^2)x_{I\!\!P}g(x_{I\!\!P}, \mu^2)T(b)
\end{eqnarray}

In the case of a nucleus the dipole cross section becomes \cite{sartre}:
\begin{eqnarray}
  \frac{{\rm d}\sigma_{q\bar q}^{(A)}}{{\rm d}^2b}
  =\frac{\pi^2}{N_C}r^2\alpha_s(\mu^2)x_{I\!\!P}g(x_{I\!\!P}, \mu^2)\sum_{i=1}^AT(|{\bf b}-{\bf b}_i|)
  \label{eq:bNonSateA}
\end{eqnarray}
and the coherent part of the bNonSat cross section can be obtained by
the average \cite{sartre}:
\begin{eqnarray}
  \left<\frac{{\rm d}\sigma_{q\bar q}^{(A)}}{{\rm d}^2b}\right>_\Omega =
  \frac{\pi^2}{N_C}r^2\alpha_s(\mu^2)x_{I\!\!P}g(x_{I\!\!P},\mu^2)AT_A(b)
  \label{eq:nosatCoherent}
\end{eqnarray}
while the second moment of the amplitude is averaged over nucleon configurations
as above.
\subsubsection{Corrections to the dipole cross section}
\label{corrections}
In the derivation of the dipole amplitude only the real part of the
$S$-matrix is taken into account.  The imaginary part of the
scattering amplitude can be included by multiplying the cross section
by a factor $(1+B^2)$, where $B$ is the ratio of the imaginary
and real parts of the scattering amplitude.  It is calculated using
\cite{Kowalski:2006hc}:
\begin{eqnarray}
    B=\tan\left(\lambda\frac{\pi}{2}\right),~{\rm where}~
    \lambda\equiv\frac{\partial\ln\left(\mathcal{A}_{T,L}^{\gamma*p\rightarrow Vp}
    \right)}   {\partial\ln(1/x)}.\label{eq:corr1}
\end{eqnarray}

In the derivation of the dipole amplitude, the gluons in the two-gluon
exchange in the interaction are assumed to carry the same momentum
fraction of the proton or nucleus. To account for the cases where they 
carry different momentum fractions, a so-called skewness correction
is applied to the cross section by multiplying it by a factor
$R_g(\lambda)$, defined by \cite{Kowalski:2006hc}:
\begin{eqnarray}
    R_g(\lambda)=\frac{2^{2\lambda+3}}{\sqrt{\pi}}\frac{\Gamma(\lambda+5/2)}{\Gamma(\lambda+4)}
\label{eq:corr2}
\end{eqnarray}
where $\lambda$ is defined in eq.~(\ref{eq:corr1}).

These corrections are important for describing HERA data.
Where the models are valid the corrections are typically around 60\% of the cross section,
with approximately 45\% attributable to the skewness correction.
The corrections grow dramatically in the large-$x$ range outside the validity of the models,
where $\xpom>10^{-2}$.
\subsection{Calculating Cross sections}
The total diffractive differential cross section is:
\begin{eqnarray}
  \frac{{\rm d}^3\sigma_{\rm total}}{{\rm d}Q^2{\rm d}W^2{\rm d}t}=
   \sum_{T, L}\frac{R_g^2(1+B^2)}{16\pi}
  \frac{{\rm d}n_{T, L}^\gamma}{{\rm d}Q^2{\rm d}W^2}
  \left<|\mathcal{A}_{T, L}|^2\right>_\Omega~~ 
  \label{eq:totalCS}
\end{eqnarray}
where ${\rm d}n_{T, L}^\gamma/{\rm d}Q^2{\rm d}W^2$ is the flux
of transversely and longitudinally polarized virtual photons,
and the average over configurations $\Omega$ is defined
in eq.~(\ref{eq:average}).
The photon flux may be that from an electron, as in the case of
$ep$ and $e$A collisions, but it may also emanate from 
protons or ions in the case of ultraperipheral collisions (UPC)
in hadron colliders, as described in e.g.~\cite{Klein:1999gv}.

The coherent part of the cross section is:
\begin{eqnarray}
  \frac{{\rm d}^3\sigma_{\rm coherent}}{{\rm d}Q^2{\rm d}W^2{\rm d}t}=
  \sum_{T, L} \frac{R_g^2(1+B^2)}{16\pi}
  \frac{{\rm d}n_{T, L}^\gamma}{{\rm d}Q^2{\rm d}W^2}
  \left|\left<\mathcal{A}_{T, L}\right>_\Omega\right|^2~~
  \label{eq:coherentCS}
\end{eqnarray}
while the incoherent part is the difference between the total
and coherent cross sections. The incoherent part directly gives the probability for the
nucleus breaking up.

For the the second moment of the amplitude,
for each nucleon configuration $\Omega_i$, one needs to calculate the integral:
\begin{eqnarray}
  \mathcal{A}_{T, L}(Q^2, \Delta, \xpom, \Omega_i)=
  \!\int r\dint r\frac{\dint z}{2}\dint^2{\bf b}
  \left(\Psi^*_V\Psi\right)_{T,L}(Q^2, r, z)
  J_0([1-z]r\Delta)
  e^{-i{\bf b}\cdot{\bf \Delta}}
  \frac{\dint\sigma_{q\bar q}}{\dint^2{\bf b}}(\xpom,r,{\bf b}, \Omega_i)
  ~~~~~&~&
  \label{eq:moment2}
\end{eqnarray}
where the dipole cross section is defined in eq.~(\ref{eq:bSateA}) for
bSat and in eq.~(\ref{eq:bNonSateA}) for bNonSat.  For $e$A, there is
no angular symmetry in ${\bf b}$ which makes this integral a complex number. We
average over 500 nucleon configurations, giving 1000 such integrals
for each point in phase-space (500 for each polarization).

For the first moment of the amplitude, the integral to calculate is:
\begin{eqnarray}
  \left<\mathcal{A}_{T,L}(Q^2, \Delta, \xpom)\right>_\Omega\!=\!
  \int\pi r\dint r\dint zb\dint b
  \left(\Psi^*_V\Psi\right)_{T, L}(Q^2, r, z)
  J_0([1-z]r\Delta)
  J_0(b\Delta)
  \left<\frac{{\rm d}\sigma_{q\bar q}}{{\rm d}^2{\bf b}}\right>_\Omega(\xpom, r, b)
  \label{eq:moment1}
\end{eqnarray}
where the average in the last term is defined in
eq.~(\ref{eq:analytical}) for bSat and in eq.~(\ref{eq:nosatCoherent})
for bNonSat.

The dipole models described here are only valid for small values of
$x_{I\!\!P}<10^{-2}$ and not too small values of $\beta\equiv x/\xpom$.  If
$\beta$ becomes too small the $q\bar q$ dipole becomes unphysically
large \cite{Kowalski:2008sa}. To rectify this one would need to
include higher Fock state dipoles, such as $q\bar q g$.
However, this growth has no 
effect on the actual production cross sections (eq.~(\ref{eq:totalCS}) and (\ref{eq:coherentCS})) 
due to the implicit cut-off of the wave-overlap at already moderate radii as discussed in \cite{sartre}.
To nevertheless protect against this unphysical behavior, we introduce a cut-off in the dipole radius 
of $r<3$ fm for protons and $r<3R_0$ for nuclei, where $R_0$ is the nucleus' 
radius given in the Woods-Saxon parametrization. We varied the cut-off in a wide range and did not
observe any changes in the resulting cross sections.

\section{Description of the program}
\label{program}

\sartre\ is not a stand-alone program but consists of a set of C++ classes and C-functions that 
together form the Application Programming Interface (API). 
Although the classes are primarily designed to provide tools to generate events they also can be used 
to calculate cross sections or study model dependencies. 

In this section we will give an overview of
most of the classes in \sartre~and how they can be used. We will give, by means of example 
programs, an overview on how to deploy them to generate amplitude look-up tables, cross sections, and events.

The master equation  of \sartre~is the total cross section described in eq.~(\ref{eq:totalCS}).
This cross sections is used as a probability density function (PDF) from which
a phase-space point $(Q^2, W^2, t)$ is drawn for each event. The phase-space
point together with the given beam energies fully determines the final state, except for the azimuthal angle of the vector meson,
which we distribute uniformly.

To determine the diffractive cross section in $e$A at a point in phase-space a complex 4-dimensional integral needs to be calculated for
500 nuclear configurations.
The fact that 1000 (500 for each polarization) such 4-dimensional integrals have to be calculated at each point in 
phase-space makes efficient event generation 
essentially impossible. The only viable approach is to compute the first and second moments of
the amplitudes separately and store the result in 3-dimensional lookup-tables in
$Q^2$, $W^2$, and $t$. This has to be done for each nuclear species, each final-state vector meson and DVCS,
each polarization, as well as for 
each dipole model (bSat or bNonSat). 
This requires a set of four look-up tables:
\begin{eqnarray} 
\left<\mathcal{A}_{T}\right>_{\Omega}(Q^2, W^2, t),~ 
\left<\mathcal{A}_{L}\right>_{\Omega}(Q^2, W^2, t),~
\left<|\mathcal{A}_{T}|^2\right>_{\Omega}(Q^2, W^2, t),~{\rm and}~
\left<|\mathcal{A}_{L}|^2\right>_{\Omega}(Q^2, W^2, t)
 \end{eqnarray}
 where $L$ and $T$ denote
 longitudinal and transverse polarization of the photon, respectively. These look-up
 tables contains all the physics of the dipole-models described in section \ref{dipolemodel}.
 
In addition we also provide tables for calculating the phenomenological corrections described in 
eqs.~(\ref{eq:corr1}) and (\ref{eq:corr2}). They hold the values of $\lambda$ defined in eq.~(\ref{eq:corr1}) for 
each $Q^2$, $W^2$ and $t$ bin matching those of the amplitude tables. We provide one table for each species of 
vector meson. However, we also put a fall back solution in place that derives $\lambda$ from the proton amplitude tables
should the lambda table not be available.

Hence, the classes in \sartre~have to accomplish two tasks. The first is the generation of the 
look-up tables for the first and second moments of the amplitude, the second is the
actual generation of the events using the look-up tables to calculate a cross section 
from which the PDF is constructed. Without further features enabled (such as nuclear break-up), 
the event generation can simulate around a million events in a matter of 3 minutes on our laptop computers. 

\begin{figure}[bt]
    \begin{center}
    \includegraphics[width=0.8\textwidth]{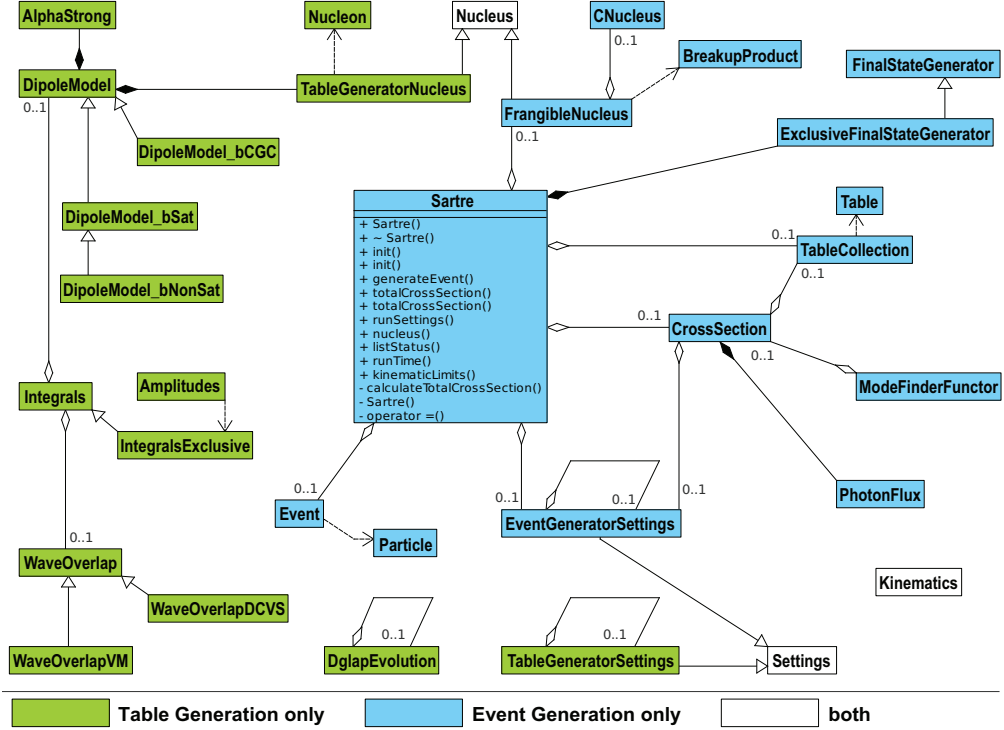}
    \end{center}
    \caption{\label{fig:scheme}
    Class Diagram of the most important classes used in \sartre. Operations and attributes are omitted for clarity, except for 
    class \texttt{Sartre} where all public operations are listed. Classes shown in green are used only for the table generation, 
    classes in blue only for event generation. We use Unified Modeling Language.}
\end{figure}

To generate events the user has to provide a main program and optionally a runcard, i.e., a text file with instructions read by \sartre, that define various parameters such as beam energies, what dipole model to use, what vector meson species to generate, the number of events, and much more. 
The overall controlling class is \texttt{Sartre} while the defined parameters are managed by \texttt{EventGeneratorSettings}, a singleton class.

Figure \ref{fig:scheme} depicts the Unified Modeling Language (UML) diagram of the most important classes and their relations.
Class \texttt{Sartre} is shown with all public operations listed while for all other classes
they are omitted for clarity. 

\begin{figure}[htb]
    \begin{center}
    \includegraphics[width=0.4\paperheight]{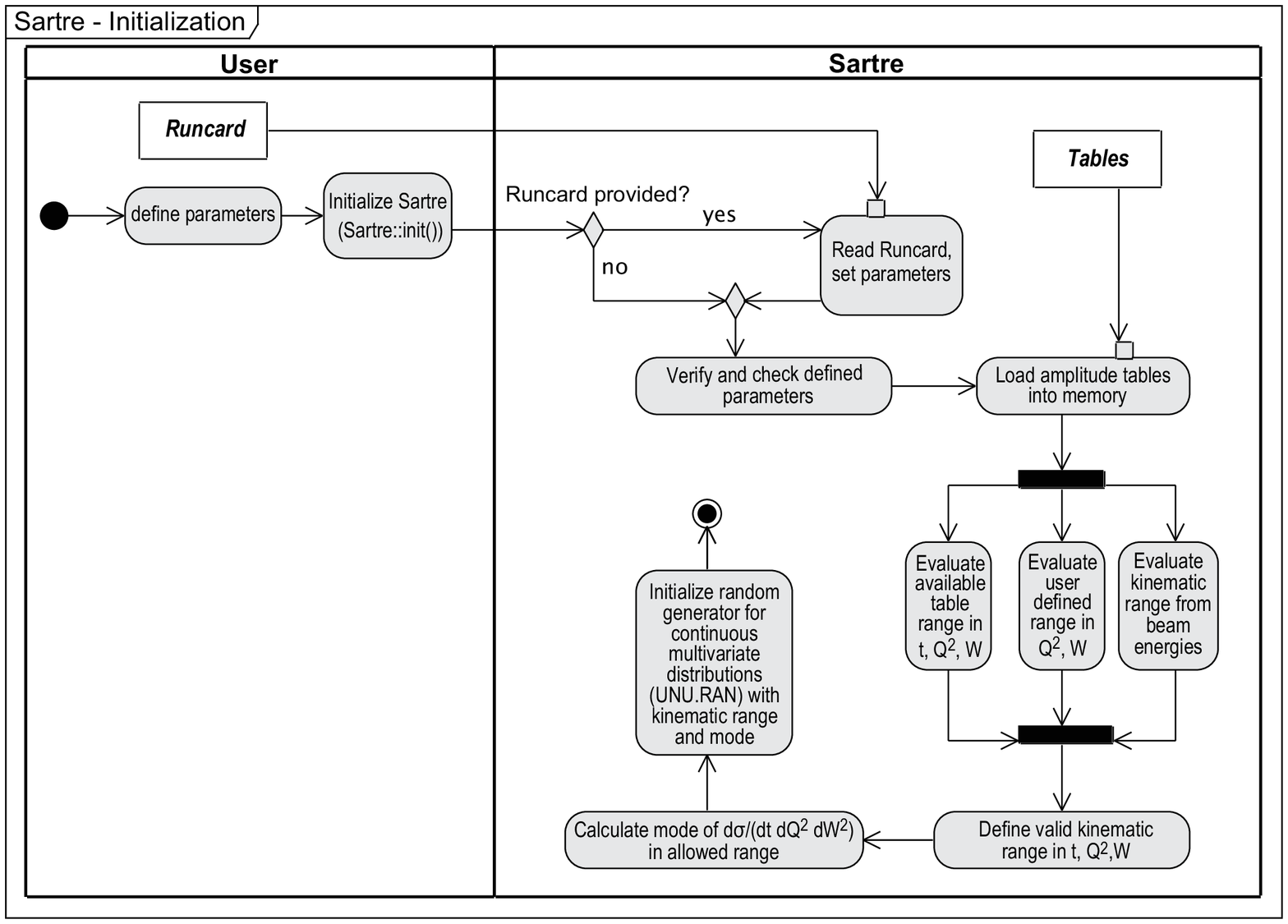}
    \end{center}
    \caption{\label{fig:init} Activity diagram showing the process for initializing \sartre~for event generation.}
\end{figure}

In the users main program the first step is to create an instance of \texttt{Sartre} and initialize all parameters.
The user can provide all parameters either programatically or through a runcard.
The UML activity diagram in Fig.~\ref{fig:init} illustrates the initialization process.
At the end, the following tasks are accomplished: (i) the amplitude look-up tables are read and stored in memory, (ii) the
actual kinematic range in $Q^2$, $W$, and $t$ is determined from the table limits, the beam energies, and the user input, and (iii) the 
random number generator (\texttt{UNU.RAN}) is initialized using a functor that computes the cross section for a given $Q^2$, $W$, and $t$, using the stored amplitude tables.
To initialize \texttt{UNU.RAN} the mode of the PDF has to be computed, which is numerically challenging since the mode typically 
lies at the border of the kinematically valid range. During the initialization the instance of \texttt{Sartre} prints the status and general information depending on the level of verbosity set by the user.
Once initialized the user can only change parameters that are related to event processing; changing parameters that are related to 
kinematic range or physics processes will have no effect.

\begin{figure}[th!]
    \begin{center}
    \includegraphics[width=0.5\paperwidth]{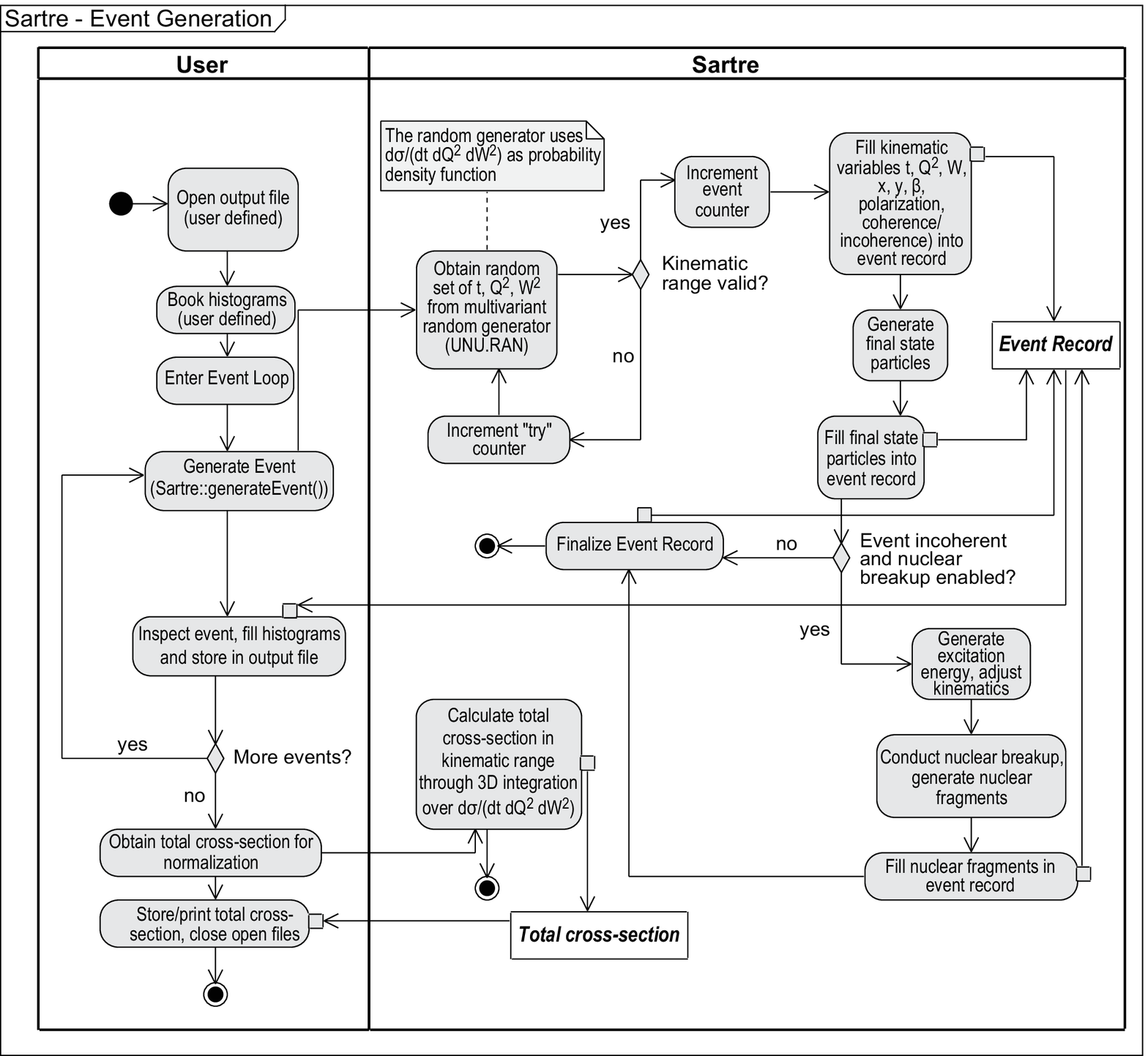}
    \end{center}
    \caption{\label{fig:generate}Activity diagram showing the process of event generation using \sartre.}
\end{figure}
\begin{figure}[bh!]
    \begin{center}
    \includegraphics[width=0.5\paperwidth]{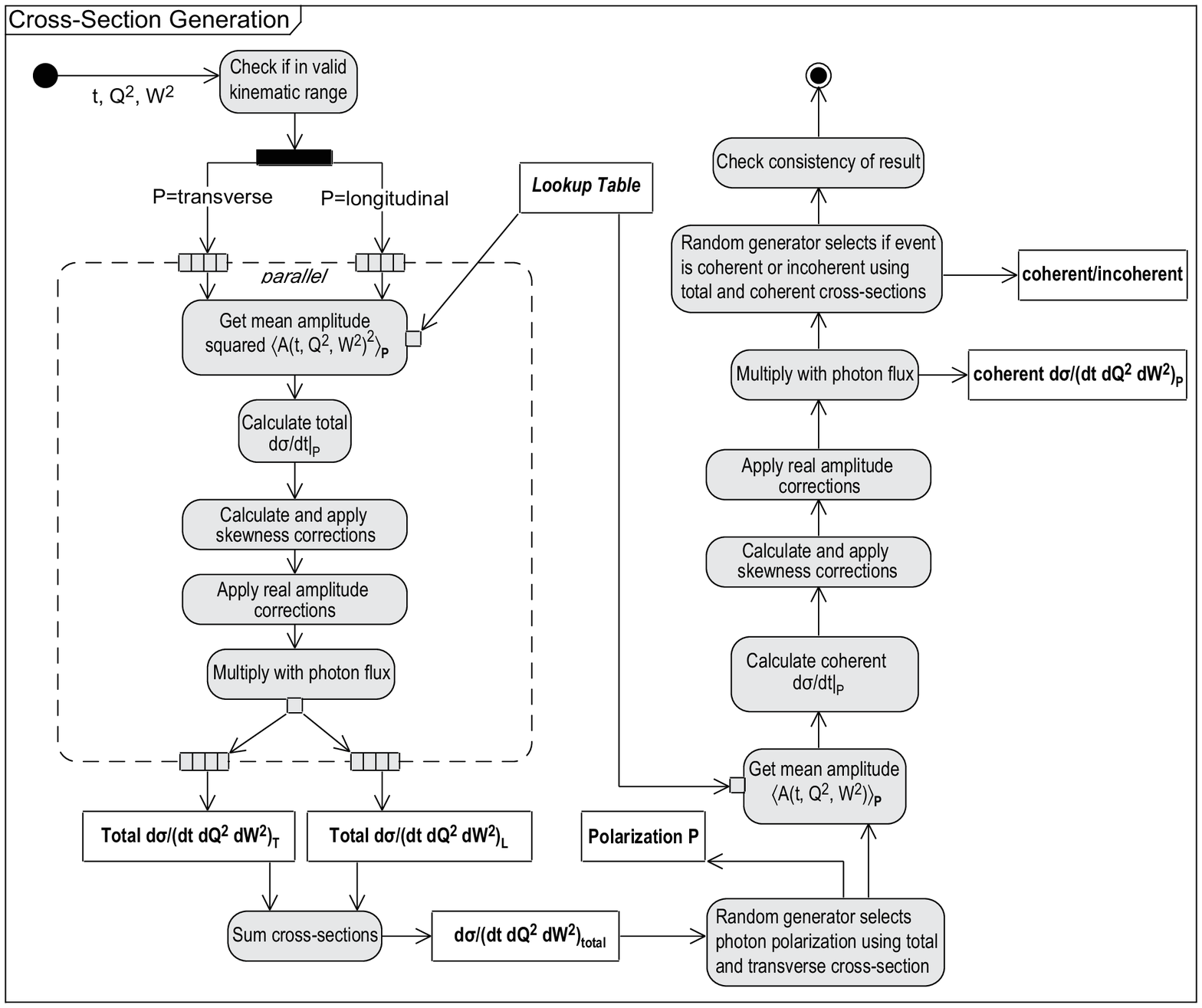}
    \end{center}
    \caption{\label{fig:cs} Activity diagram depicting the calculation of the cross section in a given
    	kinematic range.}
\end{figure}

Once \texttt{Sartre} is initialized, the initialized instance can be used for generating events.
Figure \ref{fig:generate} shows the UML activity diagram of the overall process while Figure \ref{fig:cs} depicts the
actual process of deriving the cross section(s).
From the PDF, the random generator (\texttt{UNU.RAN}) provides a phase-space point in $Q^2$, $W$, and $t$. However,
\texttt{UNU.RAN} internally operates on rectangular borders exceeding in certain areas of the phase space the limits of the kinematic acceptance.
Therefore additional kinematic checks have to be conducted. If the event is rejected, a new one is automatically generated. The difference between ``tried" and generated events is typically small ($<1\%$). The values of $Q^2$, $W$, and $t$, together with the beam energies and masses
of the incoming and outgoing particles fully define the final state which is 
calculated by the final state generator (class \texttt{ExclusiveFinalStateGenerator}), giving 4-momenta of all out-going particles.
By comparing the incoherent and coherent cross sections in the event, it is then
decided probabilistically whether the nucleus subsequently breaks up or not. 
If it does, the outgoing nucleus may then fragment by the 
\texttt{GEMINI++} intra nuclear cascade \cite{gemini++}. 
\texttt{GEMINI++} is a statistical model code which describes the nuclear
de-excitation, providing the break-up products from neutrons up to the
heaviest fragments. It needs as input the excitation energy $E^*$ of the nucleus,
which is given by:
\begin{eqnarray}
  E^*=(M_Y-m_n)\cdot A
\end{eqnarray}
We assume that the diffractive mass $M_Y$ is distributed according to:
\begin{eqnarray}
  \frac{\dint N}{\dint M_Y^2}\propto\frac{1}{M_Y^2}.
\end{eqnarray}
Note that $M_Y$ cannot be uniquely determined from kinematics alone.

Once the events have been generated one typically wants to calculate the total
cross section for the kinematic range in question in order to allow the absolute normalization 
of the generated spectra.  This can be
done independently of the event generation. 
\sartre~uses the Adaptive multi-dimensional algorithm in \texttt{ROOT} \cite{Brun:1997pa} 
and \texttt{GSL}  \cite{GSL} 
to integrate the cross section over
a valid kinematic range. If that fails, the fall-back option is VEGAS.

What follows is a description of how to obtain \sartre, and 
of the different components of the program.

\subsection{Installation}
The source-code is available at: \texttt{http://code.google.com/p/sartre-mc/}
The complete source-code including the tables and documentation can be downloaded as a tar ball
(recommended) or alternatively extracted form the subversion repository.

Unpack the downloaded tar ball: \texttt{tar -xzvf sartre-$<$version$>$.tgz}

The main directory (\texttt{sartre}) contains a \texttt{INSTALL.HTML} file with detailed instruction
how to build and set up \sartre~using the provided \texttt{cmake} files.

\sartre~is by default installed in \texttt{/usr/local/sartre} containing the libraries (\texttt{libs/}), 
the include files (\texttt{include/}), the html documentation (\texttt{docs/}), the tables (\texttt{tables/}), \texttt{GEMINI++} 
( \texttt{gemini/}),
a directory with various example programs (\texttt{examples/}), and a directory with binaries (\texttt{bin/}) that contains
tools to query and browse the amplitude tables. The installation directory can be set by the user via \texttt{cmake} command line options.

Sartre requires two additional packages to be installed: \texttt{ROOT} and
the GNU Scientific Library (\texttt{GSL}).  \texttt{ROOT} must contain
the \texttt{Unuran} and \texttt{MathMore} components. 

\subsection{Enumerations}
Table \ref{tab:Enumerations} shows a list of the few enumerations that are used throughout in \sartre.
\begin{table}[htb]
	\begin{center}
		  \begin{tabular}{|c|c|}
		  \hline
	      	Enumeration & Values \\
	       \hline\hline
	       \texttt{DipoleModelType} & \texttt{bSat, bNonSat, bCGC} \\ \hline
		\texttt{GammaPolarization} & \texttt{transverse, longitudinal} \\ \hline
		\texttt{AmplitudeMoment} & \texttt{mean\_A, mean\_A2, lambda\_A} \\\hline
		\texttt{DiffractiveMode} & \texttt{coherent, incoherent} \\
      \hline
    \end{tabular}
  \end{center}
  \caption{\label{tab:Enumerations}  A list of enumerations used throughout \sartre~to 
  	identify the dipole model in use, the polarization of the virtual photon,
	which moment of the amplitude is being used and whether the event is coherent or
	incoherent.}
\end{table}

\subsection{A description of classes}
We here give a description of those classes of the \sartre~API which we deem to
be most relevant for the user.
\subsubsection{Class Sartre}
\texttt{Sartre} is the central class of the event generation. It provides methods to 
initialize, control and run the even generator. It can also calculate cross section
in a kinematic range. At a minimum the user needs to invoke two methods:
\begin{enumerate}
	\item \verb+bool Sartre::init(char* runcard_file)+
	\item \verb+Event* Sartre::generateEvent()+
\end{enumerate}
where \texttt{runcard\_file} is the name of the text file containing the run parameters (the runcard), and \texttt{Event} is 
the class holding the event record.

\subsubsection{Class Event}
The class \texttt{Event} holds the complete event record. The user has the option to print the event record.
The event record contains the 4-momenta of all particles involved in the event, as well
as the event's kinematic variables $Q^2$, $W$, $t$, $x$, $\xpom$, $B$, $s$, $y$, the polarization
of the virtual photon, whether the event is coherent or incoherent and the number of the event.
The event record also has information on parent and daughter particle relationships. If nuclear breakup is
enabled it will also hold the nuclear fragments and their 4-momenta.

\subsubsection{Class BreakupProduct}
\texttt{BreakupProduct} holds information about the fragments created by nuclear breakup in incoherent $e$A collisions. It holds the results of \texttt{GEMINI++}, the Monte Carlo used to describe the nuclear breakup. Each fragment from \texttt{GEMINI++} is stored as a \texttt{BreakupProduct} structure and kept in a vector that is later passed and stored in the event record if nuclear breakup is switched on (via a runcard or programmatically). The emission time of the fragment is in units of $10^{-21}$ seconds since the creation of the compound nucleus in the nucleus rest frame. This unit is common in nuclear physics and we left it as provided by \texttt{GEMINI++}. Note that the 4-momenta are already boosted into the \sartre~lab frame and are not expressed per nucleon here as it is done in the main event record.
%

\subsubsection{Class Nucleus}
The Nucleus class contains the necessary information about the nucleus, such as name, radius, spin, mass as well as the referring Woods-Saxon distribution to describe its density as a function of impact parameter. This class is used in the generation of the amplitude look-up tables, while a derived class \texttt{FrangibleNucleus} is used for event generation in class Sartre. The user does not have to care about the class or its initialization since it is handled internally in \sartre. However, it is always available through a call of \texttt{Sartre::nucleus()}.

At the moment this class is able to describe the following nuclei: proton (1), oxygen (16), 
aluminum (27), calcium (40), copper (63), cadmium (110), gold (197), and lead (208).
\subsubsection{Class Kinematics}
The \texttt{Kinematics} class contains functions that calculate kinematic variables out of 
other variables. It contains functions for calculating the variables 
$x$, $y$, $s$, $W^2$, $\xpom$. It also contains the static beam particle 4-vectors
and functions that calculate the kinematically allowed limits of $Q^2$, $W^2$, and $t$.
\texttt{Kinematics} also contains a method to conduct an overall thorough check of the event kinematics.
This function is invoked in several places to guarantee the numeric integrity of the generated events.

\subsubsection{Class Particle}
The \texttt{Particle} class contains all information needed to describe any particle used in \sartre. The particles are stored in an internal list in the \texttt{Event} class. It is a lightweight class without any member functions and all data members are public. The data members hold information on the particle's status (stable or decayed), 4-momentum, parent particle(s), and daughter particle(s). Each particle is uniquely identified by an index number.

Note that momentum and energy of nuclei are always expressed as per-nucleon.
\subsubsection{Class EventGeneratorSettings}
The main API for event generation is class \texttt{Sartre}. However, \texttt{Sartre} does not handle the setup parameters at all but defers that job to the ``Settings'' classes. These classes are not only simple containers for parameters but provide lots of functionality to load, print, and manage parameters and also provide particle parameter lookup features. The basic functionality is provided by the \texttt{Settings} class, inherited by the \texttt{EventGeneratorSettings} class. For creating the amplitude table a separate class needs to be used (\texttt{TableGeneratorSettings}). \texttt{EventGeneratorSettings} only handles parameters needed for generating events and for calculating cross sections.

\texttt{EventGeneratorSettings} is a singleton class, i.e., only one instance exists at any time. One can always obtain the actual instance using the static \texttt{EventGeneratorSettings::instance()} method.

\texttt{EventGeneratorSettings} (via \texttt{Settings}) also provides the "runcard" mechanism, that is the possibility to store all parameters to run \sartre, in a text file (the runcard) and read them in.

Parameters managed by \texttt{EventGeneratorSettings} can be set and used via access functions. Each access function has an equivalent runcard name.  

\subsubsection{Class TableGeneratorSettings}
Inherits from \texttt{Settings}. Handles the settings for running the generator of look-up tables. Provides information such as
which dipole model to use, in which kinematical limits and how to bin the lookup tables.

\subsubsection{Class ExclusiveFinalStateGenerator}
Calculates the final state given the phase-space point $(t, Q^2, W^2)$ and beam energies.
The final state is stored in the event record (class \texttt{Event}).

\subsubsection{Class Amplitudes and Class Integrals}
The \texttt{Amplitudes} class calculates the quantities 
$\left<\left|\mathcal{A}^{\gamma^*p}_{L, T}\right|^2\right>_\Omega(W^2, Q^2, t)$ and
$\left<\mathcal{A}^{\gamma^*p}_{L, T}\right>_\Omega(W^2, Q^2, t)$ at a given point
in phase space. In \sartre~it is used by the table generator and the result is stored in 
look-up tables. The integrations are performed by the \texttt{Integrals} class. There is an
option to calculate the integrals for different nuclear configurations on parallel threads,
using the BOOST library \cite{BOOST}.
The integrations are using the CUBA library.
\subsubsection{Class Table}
This is a class of functions for creating a new look-up table of the amplitude momenta,
or to read information from an existing look-up table. How tables are stored and read
is described in further detail in \ref{app:table}.

We provide a program ``\texttt{tableInspector}" to inspect and browse the tables. It can be used
either for getting basic information about the table, such as binning and ranges, but
also to get detailed information about the content in each bin.

We also provide a program called ``\texttt{tableMerger}" which can merge some tables into
one combined table. This is very useful when one creates parts of a table in parallel
on different computing nodes and then merge them into one table. For \texttt{tableMerger}
to work, the granularities of the tables have to match, as well as the ranges, in two
out of the three dimensions. 

There are also tables which stores the values of $\lambda$ described in eq.~(\ref{eq:corr1}). These
tables are used by \sartre~to calculate the real part and skewness corrections to the cross sections.
\subsubsection{Class TableCollection}
Several look-up tables may exist for a certain process. These tables may cover different
kinematic regions and have different binning. The tables may also overlap in some
kinematic regions. \sartre~calculates the cross sections from 
\texttt{TableCollection} rather than \texttt{Table}. 
When \sartre~calls \texttt{TableCollection}, it chooses
from which of these tables to extract the information.

\section{Example programs and runcards}
In this sections we provide examples of user programs making use of the classes in \sartre.
To generate events with \sartre, one first need lookup tables. There is a collection of such tables included in the package.
Here we provide an example of how to use the event generator in \sartre, as well as providing an example 
program for how to generate the lookup tables. 

The normal user is not expected to do the latter. Before generating lookup tables, the user need to 
produce a file containing nuclear configurations, which is produced with the  
example program \texttt{createBSatBDependenceTable}, 
and put it in a directory as given in the runcard. 
\label{examples}
\subsection{Runcard for event generation}
The following is an example of a runcard ``\texttt{sartreRuncard.txt}'' for generating $J/\psi$ mesons in electron-gold collisions:

\texttt{
\begin{lstlisting}
//===================================================================
// 
//  Example Runcard for Sartre Event Generator. 
// 
//  Comments start with a # or // 
//  Name and value are separated by a "=":  name = value 
//  (alternatively ":" can be used as well) 
//===================================================================
# 
#  Define beams [GeV]
# 
eBeamEnergy =  20 
hBeamEnergy =  100   

# Define atom number of nucleus
A           =  197 
 
#  Number of events and printout frequency 
numberOfEvents =  1000
timesToShow    =  20 
 
#  Set verbosity 
verbose       =  false
verboseLevel  =  1 
 
#  Rootfile  or alternative file to store the generated events
rootfile =  example.root 
 
#  Model parameters (jpsi=443, phi=333, rho=113, DVCS=22)
vectorMesonId =  443   
dipoleModel =  bSat 
 
#  Kinematic range min > max means no limits (given by table range) 
#  Q2 in GeV2, W in GeV
Q2min =  1 
Q2max =  10 
Wmin  =  10 
Wmax  =  60 
 
#  Corrections 
correctForRealAmplitude =  true 
correctSkewedness       =  true
#maxLambdaUsedInCorrections = 0.6

# Ultra Peripheral Collisions
# Simulate UPC?
UPC = false
# Atom number of photon emitting nucleus:
UPCA = 197

#  Use Gemini++ to evaporate nucleus in incoherent event?
enableNuclearBreakup = true 
maxNuclearExcitationEnergy = 0.5
 
#  Random generator seed (if not given current time is used) 
#seed =  2011987 
\end{lstlisting}}

\subsection{Generating events}
\label{exampleEvents}
The following is an example of a user program ``\texttt{sartreMain}'' for generating events with \sartre, using a runcard for
user input:
\texttt{
\begin{lstlisting}
#include <iostream>   
#include <cmath>   
#include "TTree.h"   
#include "TFile.h"   
#include "Sartre.h"   
            
using namespace std;   
   
struct rootSartreEvent {
    double t;   
    double Q2;   
    double x;   
    double s;   
    double y;   
    double W;   
    double xpom;   
    int    iEvent;   
    int    pol;      // 0=transverse or 1=longitudinal   
    int    dmode;    // 0=coherent, 1=Incoherent   
};   
   
rootSartreEvent myRootSartreEvent;

int main(int argc, char *argv[])   
{       
    //   
    //  Check command line arguments   
    //   
    if (! (argc == 2 || argc == 3) ) {   
        cout << "Usage:  " << argv[0] << " runcard [rootfile]" << endl;     
        return 2;   
    }   
    string runcard = argv[1];

    //   
    //  Create the generator and initialize it.    
    //  Once initialized you cannot (should not) change   
    //  the settings w/o re-initialing sartre.   
    //   
    Sartre sartre;   
    bool ok = sartre.init(runcard);   
    if (!ok) {   
        cerr << "Initialization of sartre failed." << endl;   
        return 1;   
    }   
    EventGeneratorSettings* settings = sartre.runSettings();   
    settings->list();  
 
    //   
    //  ROOT file    
    //   
    string rootfile;   
    if (argc == 3)    
        rootfile = argv[2];   
    else   
        rootfile = settings->rootfile();   
       
    TFile *hfile = 0;   
    if (rootfile.size()) {   
        hfile  = new TFile(rootfile.c_str(),"RECREATE");   
        cout << "ROOT file is '" <<  rootfile.c_str() << "'." << endl;   
    }   
   
    //   
    //  Setup ROOT tree    
    //   
    TLorentzVector *eIn = new TLorentzVector;   
    TLorentzVector *pIn = new TLorentzVector;   
    TLorentzVector *vm = new TLorentzVector;   
    TLorentzVector *eOut = new TLorentzVector;   
    TLorentzVector *pOut = new TLorentzVector;   
    TLorentzVector *gamma = new TLorentzVector;   
    TTree tree("tree","sartre");   
    tree.Branch("event", &myRootSartreEvent.t,    
                "t/D:Q2/D:x/D:s/D:y/D:W/D:xpom/D:
		 iEvent/I:pol/I:dmode/I");   
    tree.Branch("eIn",  "TLorentzVector", &eIn, 32000, 0);   
    tree.Branch("pIn",  "TLorentzVector", &pIn, 32000, 0);   
    tree.Branch("vm",   "TLorentzVector", &vm, 32000, 0);   
    tree.Branch("eOut", "TLorentzVector", &eOut, 32000, 0);   
    tree.Branch("pOut", "TLorentzVector", &pOut, 32000, 0);   
    tree.Branch("gamma","TLorentzVector", &gamma, 32000, 0);   

    //   
    //  Event generation   
    //          
    int nPrint;   
    if (settings->timesToShow())   
        nPrint = settings->numberOfEvents()/settings->timesToShow();   
    else    
        nPrint = 0;   
       
    unsigned long maxEvents = settings->numberOfEvents();   
   
    cout << "Generating " << maxEvents << " events." << endl << endl;   
   
    for (unsigned long iEvent = 0; iEvent < maxEvents; iEvent++) {   
        //   
        //  Generate one event   
        //   
        Event *event = sartre.generateEvent();   
        if (nPrint && iEvent%nPrint == 0 && iEvent != 0) {   
            cout << "processed " << iEvent << " events" << endl;   
        }   
   
        //   
        //  Print out (here only for the first few events)   
        //   
        if (iEvent < 4) event->list(); 
           
        //   
        //  Fill ROOT tree   
        //   
        myRootSartreEvent.iEvent = event->eventNumber;   
        myRootSartreEvent.t = event->t;   
        myRootSartreEvent.Q2 = event->Q2;   
        myRootSartreEvent.x = event->x;   
        myRootSartreEvent.y = event->y;   
        myRootSartreEvent.s = event->s;   
        myRootSartreEvent.W = event->W;   
        myRootSartreEvent.xpom = event->xpom;   
        myRootSartreEvent.pol = 
		event->polarization == transverse ? 0 : 1;   
        myRootSartreEvent.dmode = 
		event->diffractiveMode == coherent ? 0 : 1;   
        eIn     = &event->particles[0].p;   
        pIn     = &event->particles[1].p;   
        eOut    = &event->particles[2].p;   
        pOut    = &event->particles[6].p;   
        vm      = &event->particles[4].p;   
        gamma   = &event->particles[3].p;   
        
        tree.Fill();    
    }   
    //   
    //  That's it, finish up   
    //   
    double runTime = sartre.runTime();
    hfile->Write();     
    cout << rootfile.c_str() << " written." << endl;   
    cout << "Total cross-section: " << sartre.totalCrossSection() 
         << " nb" << endl;   
    sartre.listStatus();   
    cout << "CPU Time/event: " << 1000*runTime/maxEvents 
         << " msec/evt" << endl;   
       
    return 0;   
} 
\end{lstlisting}}
\subsection{Output and Result from Event Generation}
The following is the output resulting from running the example program ``\texttt{sartreMain}'' described in section 
\ref{exampleEvents}:
{\footnotesize 
\begin{verbatim}
/========================================================================\
|                                                                        |
|  Sartre, Version 1.10                                                  |
|                                                                        |
|  An event generator for exclusive diffractive vector meson production  |
|  in ep and eA collisions based on the dipole model.                    |
|                                                                        |
|  Copyright (C) 2010-2013 Tobias Toll and Thomas Ullrich                |
|                                                                        |
|  This program is free software: you can redistribute it and/or modify  |
|  it under the terms of the GNU General Public License as published by  |
|  the Free Software Foundation, either version 3 of the License, or     |
|  any later version.                                                    |
|                                                                        |
|  Code compiled on Aug 15 2013 09:56:29                                 |
|  Run started at Wed Dec 11 15:58:28 2013                               |
\========================================================================/
Runcard is '/Users/tollto/sartre/paper-program/version5/generatorRuncardExample.txt'.
Hadron beam species: Au (197)
Hadron beam:   0	0	99.9956	100	(0.93827)
Electron beam: 0	0	-20	20	(0.000510999)
Dipole model: bSat
Process is e + Au -> e' + Au' + J/psi
Random generator seed: 1386795508
Sartre is initialized.


Run Settings:
                     userInt	0
                  userDouble	0
                  userString	
                 eBeamEnergy	20
                 hBeamEnergy	100
                           A	197
              numberOfEvents	1000
                 timesToShow	20
               vectorMesonId	443
                 dipoleModel	bSat
                       Q2min	1
                       Q2max	10
                        Wmin	10
                        Wmax	60
     correctForRealAmplitude	true
           correctSkewedness	true
        enableNuclearBreakup	true
  maxLambdaUsedInCorrections	0.65
  maxNuclearExcitationEnergy	0.5
             applyPhotonFlux	true
                     verbose	false
                verboseLevel	0
                         UPC	false
                        UPCA	197
                    rootfile	example.root
                        seed	1386795508

ROOT file is 'example.root'.
Generating 1000 events.

evt = 0           Q2 = 1.573         x = 1.104e-03
                   W = 37.734        y = 0.178
                   t = -0.007     xpom = 7.841e-03
                 pol = T          diff = coherent

   #          id   name       status    parents     daughters        px       py       pz        E           m
   0          11   e-              4    -     -      2     3      0.000    0.000  -20.000   20.000   5.110e-04 
   1  1000791970   Au(197)         4    -     -      6     -      0.000    0.000   99.996  100.000       0.938 
   2          11   e-              1    0     -      -     -      0.861   -0.743  -16.419   16.458   5.110e-04 
   3          22   gamma           2    0     -      4     5     -0.861    0.743   -3.581    3.542      -1.254 
   4         443   J/psi           1    3     -      -     -     -0.826    0.669   -2.809    4.314       3.097 
   5         990   pomeron         2    3     3      6     -     -0.034    0.073   -0.772   -0.772      -0.082 
   6  1000791970   Au(197)         1    1     5      -     -     -0.034    0.073   99.223   99.228       0.938 

evt = 1           Q2 = 3.684         x = 2.582e-03
                   W = 37.734        y = 0.178
                   t = -0.007     xpom = 9.309e-03
                 pol = L          diff = coherent

   #          id   name       status    parents     daughters        px       py       pz        E           m
   0          11   e-              4    -     -      2     3      0.000    0.000  -20.000   20.000   5.110e-04 
   1  1000791970   Au(197)         4    -     -      6     -      0.000    0.000   99.996  100.000       0.938 
   2          11   e-              1    0     -      -     -      1.679    0.455  -16.387   16.479   5.110e-04 
   3          22   gamma           2    0     -      4     5     -1.679   -0.455   -3.613    3.521      -1.919 
   4         443   J/psi           1    3     -      -     -     -1.604   -0.426   -2.701    4.432       3.097 
   5         990   pomeron         2    3     3      6     -     -0.075   -0.029   -0.911   -0.911      -0.082 
   6  1000791970   Au(197)         1    1     5      -     -     -0.075   -0.029   99.084   99.089       0.938 

evt = 2           Q2 = 3.684         x = 3.342e-03
                   W = 33.156        y = 0.138
                   t = -0.007     xpom = 1.205e-02
                 pol = T          diff = coherent

   #          id   name       status    parents     daughters        px       py       pz        E           m
   0          11   e-              4    -     -      2     3      0.000    0.000  -20.000   20.000   5.110e-04 
   1  1000791970   Au(197)         4    -     -      6     -      0.000    0.000   99.996  100.000       0.938 
   2          11   e-              1    0     -      -     -     -1.634   -0.711  -17.199   17.291   5.110e-04 
   3          22   gamma           2    0     -      4     5      1.634    0.711   -2.801    2.709      -1.919 
   4         443   J/psi           1    3     -      -     -      1.563    0.749   -1.612    3.898       3.097 
   5         990   pomeron         2    3     3      6     -      0.071   -0.038   -1.189   -1.189      -0.082 
   6  1000791970   Au(197)         1    1     5      -     -      0.071   -0.038   98.807   98.811       0.938 

evt = 3           Q2 = 2.712         x = 2.906e-03
                   W = 30.516        y = 0.117
                   t = -0.005     xpom = 1.319e-02
                 pol = T          diff = coherent

   #          id   name       status    parents     daughters        px       py       pz        E           m
   0          11   e-              4    -     -      2     3      0.000    0.000  -20.000   20.000   5.110e-04 
   1  1000791970   Au(197)         4    -     -      6     -      0.000    0.000   99.996  100.000       0.938 
   2          11   e-              1    0     -      -     -      1.173    1.009  -17.633   17.701   5.110e-04 
   3          22   gamma           2    0     -      4     5     -1.173   -1.009   -2.367    2.299      -1.647 
   4         443   J/psi           1    3     -      -     -     -1.244   -1.010   -1.029    3.636       3.097 
   5         990   pomeron         2    3     3      6     -      0.071    0.001   -1.337   -1.337      -0.073 
   6  1000791970   Au(197)         1    1     5      -     -      0.071    0.001   98.658   98.663       0.938 

processed 50 events
processed 100 events
processed 150 events
processed 200 events
processed 250 events
processed 300 events
processed 350 events
processed 400 events
processed 450 events
processed 500 events
processed 550 events
processed 600 events
processed 650 events
processed 700 events
processed 750 events
processed 800 events
processed 850 events
processed 900 events
processed 950 events
processed 1000 events
All events processed

example.root written.
Total cross-section: 94.4 nb
Event summary: 1000 events generated, 1000 tried
Total time used: 0 min 15 sec
CPU Time/event: 15 msec/evt
\end{verbatim}}

\begin{figure}
   \begin{center}
    \includegraphics[width=0.5\paperwidth]{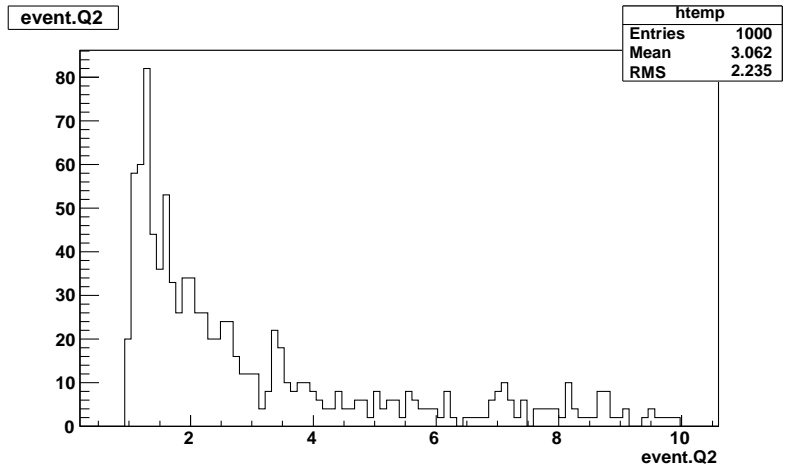}
    \caption{\label{fig:Q2}The resulting distribution of $Q^2$ from running the example program described in the text.}
    \end{center}
\end{figure}

The produced file ``\texttt{example.root}" includes a \texttt{ROOT} tree, including the information of the 
each event, including event number, $Q^2$, $x$, $W$, $t$, etc. 
The tree also contains the four-vectors of each produced particle in the event, as well as the beam-particles.
In Fig.~\ref{fig:Q2} is shown the resulting $Q^2$ distribution from running the example program.

\subsection{Runcard for look-up table generation}
The following is showing an example runcard for the table generator, named ``\texttt{tableGeneratorRuncard.txt}'':
\texttt{
\begin{lstlisting}
//==============================================================
// 
//  Example Runcard for Sartre Table Generator. 
// 
//  Comments start with a # or #  
//  Name and value are separated by a "=":  name = value 
//  (alternatively ":" can be used as well) 
//==============================================================
# Nucleus' atomic number 
A           =  197
 
# Which moments of the amplitude to calculate: 
# 0 <A> analytical, <A2> averaged over configurations (Default)
# 1 <A> only, analytical 
# 2 <A2>, and <A> both averaged over configurations  
modesToCalculate = 0 
 
#  Rootfile This is a prefix table type ("T", "L",  "T2" or "L2"),
#  as well as ".root" will be appended  
rootfile =  bintests 
 
# Meson to generate (jpsi=443, phi=333, rho=113, DVCS=22)
vectorMesonId =  443

#  Model, options are bSat, bNonSat, bCGC  
dipoleModel =  bSat 
 
# Path for the b-dependence lookup table for bSat/bNonSat. 
# Not used with bCGC. 
bSatLookupPath = /usr/bin/sartre/bSatLookupTable/
 
numberOfConfigurations = 500
 
#  Kinematic range, Q2 and t in Gev2, W in GeV   
Q2min =  0.1 
Q2max =  10. 
Wmin  =  60 
Wmax  =  110 
tmin  =  -.5 
tmax  =  -.0 
  
# Number of bins in the tables: 
Q2bins = 30 
W2bins = 25 
tbins = 40 
 
# If a run fails, use the backup and continue: 
# If in doubt use the tableInspector to make sure the rest  
# of the runcard is as it should be.  
useBackupFile = false 
startingBinFromBackup=2 
\end{lstlisting}}

\subsection{Generating look-up tables for amplitudes}
The following is an example of a user program ``\texttt{tableGeneratorMain}'' for generating the first and second moments
of the amplitude and store them in a look-up table:
\texttt{
\begin{lstlisting}
#include <sstream>  
#include <cstdlib>  
#include <iostream>  
#include "Amplitudes.h"  
#include "TH1D.h"  
#include "TFile.h"  
#include "Constants.h"  
#include "Table.h"  
#include "TableGeneratorSettings.h"  
#include "Enumerations.h"  

int main(int argc, char *argv[]){  
    TH1::AddDirectory(false);  
    TableGeneratorSettings* settings = TableGeneratorSettings::instance();
    //  
    //  Check arguments  
    //  
    char* runcard;  
    if (argc != 4) {  
        cout<<"Usage: tableGeneratorMain runcard startBin endBin"<<endl;  
        return 1;  
    }  
    else{  
        runcard = argv[1];  
        settings->setStartBin(atoi(argv[2]));  
        settings->setEndBin(atoi(argv[3]));  
    }  
    
    settings->readSettingsFromFile(runcard);  
    settings->consolidateSettings();  
    
    int nBinQ2 =       settings->Q2bins();  
    int nBinW2 =       settings->W2bins();  
    int nBinT=         settings->tbins();  
    double Q2min=      settings->Q2min();  
    double Q2max=      settings->Q2max();  
    double Wmin=       settings->Wmin();  
    double Wmax=       settings->Wmax();  
    double W2min=Wmin*Wmin;  
    double W2max=Wmax*Wmax;  
    double tmin=       settings->tmin();  
    double tmax=       settings->tmax();  
    unsigned int massA=settings->A();  
    int vmPDG =        settings->vectorMesonId();  
    DipoleModelType model = settings->dipoleModel();  
    int startingBin=settings->startBin();  
    int endingBin=settings->endBin();  
    int modes=settings->modesToCalculate();  
    
    Table tableT;  
    Table tableL;  
    Table tableT2;  
    Table tableL2;  
    bool logQ2=true, logW2=false, logT=false, logC=true;  
    
    //set filenames for the tables:  
    string rootfile=settings->rootfile();  
    ostringstream filenameT, filenameL, filenameT2, filenameL2;  
    filenameT.str("");  
    filenameT<<rootfile<<"_bin"<<startingBin<<"_T.root";  
    filenameL.str("");  
    filenameL<<rootfile<<"_bin"<<startingBin<<"_L.root";  
    filenameT2.str("");  
    filenameT2<<rootfile<<"_bin"<<startingBin<<"_T2.root";  
    filenameL2.str("");  
    filenameL2<<rootfile<<"_bin"<<startingBin<<"_L2.root";  
    (void) tableT.create(nBinQ2, Q2min, Q2max,  
                         nBinW2, W2min, W2max,  
                         nBinT,  tmin,  tmax,  
                         logQ2, logW2, logT, logC,       // all bools  
                         mean_A, transverse,  
                         massA, vmPDG, model,  
                         filenameT.str().c_str());  
    (void) tableL.create(nBinQ2, Q2min, Q2max,  
                         nBinW2, W2min, W2max,  
                         nBinT,  tmin,  tmax,  
                         logQ2, logW2, logT, logC,      // all bools  
                         mean_A, longitudinal,  
                         massA, vmPDG, model,  
                         filenameL.str().c_str());  
    (void) tableT2.create(nBinQ2, Q2min, Q2max,  
                          nBinW2, W2min, W2max,  
                          nBinT,  tmin,  tmax,  
                          logQ2, logW2, logT, logC,      // all bools  
                          mean_A2, transverse,  
                          massA, vmPDG, model,  
                          filenameT2.str().c_str());  
    (void) tableL2.create(nBinQ2, Q2min, Q2max,  
                          nBinW2, W2min, W2max,  
                          nBinT,  tmin,  tmax,  
                          logQ2, logW2, logT, logC,      // all bools  
                          mean_A2, longitudinal,  
                          massA, vmPDG, model,  
                          filenameL2.str().c_str());  
    
    //Create and initialize the amplitudes calculator:  
    Amplitudes amps;  
    
    //Generate the the nucleon configurations:  
    amps.generateConfigurations();  
    
    // Print out settings:  
    cout<<"Tables will be generated for:"<<endl;  
    cout<<"Mode to calculate: "<<modes<<endl;  
    cout<<"Vector Meson Id: "<<vmPDG<<endl;  
    cout<<"Bins: "<<startingBin<<"-"<<endingBin<<endl;  
    cout<<"Q2 range: ["<<Q2min<<", "<<Q2max<<"], "<<nBinQ2
    <<" bins."<<endl;  
    cout<<"W2 range: ["<<W2min<<", "<<W2max<<"], "<<nBinW2
    <<" bins."<<endl;  
    cout<<" t range: ["<<tmin<<", "<<tmax<<"], "<<nBinT<<" bins."<<endl;  
    cout<<"Dipole Model: "<<model<<endl;  
    cout<<endl;  
    cout<<"Filling tables..."<<endl;  
    time_t tableFillStart = time(0);  
    for(int i=startingBin; i<endingBin; i++) {  
        double Q2, W2, t;  
        tableT.binCenter(i, Q2, W2, t);  
        //calculate contents and fill tables:  
        amps.calculate(t, Q2, W2);  
        double aT=amps.amplitudeT();  
        double aL=amps.amplitudeL();  
        if(massA>1){  
            tableT.fill(i, aT);  
            tableL.fill(i, aL);  
        }  
        if(modes!=1){  
            double aT2=amps.amplitudeT2();  
            double aL2=amps.amplitudeL2();  
            tableT2.fill(i, aT2);  
            tableL2.fill(i, aL2);  
        }  
    }  
    cout<<"CPU Time/Entry: " << double(time(0)-tableFillStart)/endingBin
    <<" s/entry."<<endl;  
    cout<<"Total time: "<<double(time(0)-tableFillStart)/60./60.
    <<" h"<<endl;  
    
    if(massA>1){  
        tableT.write();  
        tableL.write();  
    }  
    if(modes!=1){  
        tableT2.write();  
        tableL2.write();  
    }  
    
    return 0;  
}  
\end{lstlisting}}

\section{Acknowledgements}
The authors would like to thank Henri Kowalski, Tuomas Lappi, Thomas Burton and Raju Venugopalan for their input and help, and the Open Science Grid consortium
for providing resources and support. 
This work was supported by the U.S. Department of Energy under Grant
No. DE-AC02-98CH10886.


\appendix

\section{The tables}
\label{app:table}
A table in \sartre~is internally stored in a three-dimensional \texttt{ROOT} histogram. The {\texttt Table} class encapsulates the histogram and provides 
many methods to easily store and access information. The tables are stored in ROOT files. 
To create a table, one calls:
\begin{verbatim}
Table::create(int nbinsQ2, double Q2min, double Q2max,    
              int nbinsW2, double W2min, double W2max,    
              int nbinsT, double tmin, double tmax,    
              bool logQ2, bool logW2, bool logt, bool logContent,    
              AmplitudeMoment mom, GammaPolarization pol,     
              unsigned int A, int vm,    
              DipoleModelType model, const char* filename)
\end{verbatim}
where the first nine arguments define the limits in $Q^2$, $W^2$, and $t$, as well as the granularity of the table, 
\textit{i.e.}~the number of bins in each directions. The booleans, \textit{e.g.}~\texttt{logQ2}, indicate whether the binning is
linear in $Q^2$ or in $\log Q^2$, \texttt{logContent} indicates whether the content
is stored linearly or logarithmically. Also, one needs to provide information on whether
one calculates the first or second moment of the amplitude, with which polarization of the virtual
photon (see table \ref{tab:Enumerations}), with which atomic number of the nucleus and which dipole model is used. 
\texttt{Table::create} also requires a file-name to which the table can be saved. 
There is also a mechanism for making back-ups for tables during the
generation process.

All of this information (except file name) is coded into a 64bit word that is also used as the histogram title.
It is to be
interpreted as an \texttt{uitn64\_t} with the bits set as follow:
\begin{quote}
  bit 0:     content type: 0 for $\left<\mathcal{A}\right>$, 1 for $\left<|\mathcal{A}|^2\right>$    \newline
  bit 1:     polarization: $L$ for 0, $T$ for 1      \newline
  bit 2:     $t$ encoding: 0 for $|t|$, 1 for $\log(|t|)$      \newline
  bit 3:     $W^2$ encoding: 0 for linear, 1 for logarithmic      \newline
  bit 4:     $Q^2$ encoding: 0 for linear, 1 for logarithmic       \newline
  bit 5-7:   dipole model type       \newline
  bit 8-15:  mass number A      \newline
  bit 16-31: vector meson ID (PDG)      \newline
  bit 32:    content encoding: 0 in linear, 1 in logarithmic     \newline
  bit 33:    content type is $\lambda$ (bit $0=0$ in this case) \newline
  bit 33-63: not used   
\end{quote}

When reading a value from a table, e.g. to calculate a cross section in a phase-space point,
one calls the function \texttt{Table::get(double Q2, double W2, double t)}. In general, the values
of $Q^2$, $W^2$, and $t$ asked for will not correspond with a bin-center in the histogram, 
so the stored points are used to interpolate to the value asked for. For this interpolation to be
accurate, the histogram has to be fine enough in its binning. 

For the corrections, described in eq.~(\ref{eq:corr1}) and  eq.~(\ref{eq:corr2}), special tables are generated
containing $\lambda$ as described in  eq.~(\ref{eq:corr1}). If the range of the $\lambda$-table is smaller
than for the amplitude table, the fall-back option is to calculate the derivative using the proton amplitude tables.
If the range of these tables also are too small corrections will not be applied. 
%


\bibliographystyle{model1-num-names}
\bibliography{<your-bib-database>}

\end{document}